\documentclass[aps,prb,twocolumn,superscriptaddress, fleqn, showpacs]{revtex4}

\usepackage{times,xspace}
\usepackage{amsbsy,amssymb,amsmath,bm}
\usepackage{graphicx,color,rotate}
\usepackage{longtable}
\usepackage{psfrag}

\def\bbbc{{\mathchoice {\setbox0=\hbox{$\displaystyle\rm C$}\hbox{\hbox
to0pt{\kern0.4\wd0\vrule height0.9\ht0\hss}\box0}}
{\setbox0=\hbox{$\textstyle\rm C$}\hbox{\hbox
to0pt{\kern0.4\wd0\vrule height0.9\ht0\hss}\box0}}
{\setbox0=\hbox{$\scriptstyle\rm C$}\hbox{\hbox
to0pt{\kern0.4\wd0\vrule height0.9\ht0\hss}\box0}}
{\setbox0=\hbox{$\scriptscriptstyle\rm C$}\hbox{\hbox
to0pt{\kern0.4\wd0\vrule height0.9\ht0\hss}\box0}}}}

\newcommand{\ybal}{{$\beta$-YbAlB$_4$}} 
\newcommand{\aybal}{{$\alpha$-YbAlB$_4$}}
\newcommand{\gybal}{{$\gamma$-YbAlB$_4$}}
\newcommand{\yrs}{{YbRh$_2$Si$_2$}}
\newcommand{\yb}{{YbB$_2$}}

\begin{document}

\title{The Role of Crystal Symmetry in the Magnetic Instabilities of $\beta$-YbAlB$_4$ and \aybal} 

\author{D. A. Tompsett}
\affiliation{Cavendish Laboratory, University of Cambridge, Madingley Road,
  Cambridge CB3 0HE, UK}
\author{Z. P. Yin}
\affiliation{
Department of Physics, University of California, Davis, California 95616, USA
}%
\author{G. G. Lonzarich}
\affiliation{Cavendish Laboratory, University of Cambridge, Madingley Road,
  Cambridge CB3 0HE, UK}
\author{W. E. Pickett}
\affiliation{
Department of Physics, University of California, Davis, California 95616, USA
}%

\date{\today}
\begin{abstract}

Density functional theory methods are applied to investigate the properties of the new superconductor
$\beta$-YbAlB$_4$ and its polymorph $\alpha$-YbAlB$_4$. We utilize the generalized gradient approximation + Hubbard U (GGA+U) approach with spin-orbit(SO) 
coupling to approximate the effects of the strong correlations due to the open $4f$ shell of Yb. 
We examine closely the differences in crystal bonding and symmetry of $\beta$-YbAlB$_4$ and $\alpha$-YbAlB$_4$. The in-plane bonding 
structure amongst the dominant itinerant electrons in the boron sheets is shown to differ significantly.
Our calculations indicate that, in both polymorphs, the localized 4$f$ electrons
hybridize strongly with the conduction sea when compared to the related materials YbRh$_{2}$Si$_{2}$ and YbB$_{2}$.
Comparing $\beta$-YbAlB$_4$ to the electronic structure of related crystal structures indicates a key role of the 
7-member boron coordination of the Yb ion in $\beta$-YbAlB$_4$ in producing its enhanced Kondo scale and superconductivity.
The Kondo scale is shown to depend strongly on the angle between the B neighbors and the Yb ion, relative
to the $x-y$ plane, which relates some of the physical behavior to structural characteristics.
\end{abstract}


\maketitle

\begin{center}

\textbf{I. Introduction}
\end{center}
Heavy electron systems have provided key
insights into emergent quantum mechanical behavior in correlated materials.
Such systems may be tuned by the application of pressure or a magnetic field through a quantum critical point\cite{GegenwartYb1,LohneysenYb1}. In this region they may exhibit non-Fermi liquid properties\cite{GegenwartYb1, LohneysenYb1, ColemanYb1, StewartYb1} and often superconductivity\cite{MathurYb1, MonthouxYb1}. Quantum criticality has been widely reported in Ce, U and Yb systems\cite{MathurYb1, SteglichYb1, JaccardYb1, PetrovicYb1, OttYb1, StewartYb2, SchlabitzYb1}. However, while associated heavy-fermion superconductivity has been reported in Ce and U compounds\cite{SteglichYb1, JaccardYb1, PetrovicYb1, OttYb1, StewartYb2, SchlabitzYb1}, examples in Yb systems have been lacking.
The discovery of superconductivity alongside non-Fermi liquid behavior\cite{NakatsujiYb1} in the clean Yb based heavy
fermion (HF) compound \ybal~has therefore excited much interest. Thus there exists a clear impetus to
determine the electronic structure of \ybal~and its relationship to other materials.

The superconductivity and pronounced non-Fermi liquid behavior of \ybal~occur at ambient pressure\cite{NakatsujiYb1} 
and zero applied magnetic field. At zero field the resistivity $\rho\propto T^{3/2}$ (T is the absolute temperature)
and there is a logarithmic divergence in the specific heat. The introduction
of an external magnetic field tunes the system away from quantum criticality and towards conventional Fermi liquid
behavior $\rho\propto T^{2}$. 
Recent quantum oscillation studies
demonstrate heavy fermion behavior in the material\cite{O'FarrellYb1}. Furthermore, the Curie-like susceptibility at high
temperatures suggests the presence of localized moments that are pivotal to the formation of the
heavy Fermi liquid.

\ybal~is similar to its polymorph, \aybal, in exhibiting Curie-Weiss type behavior in the susceptibility at high temperature and in the formation of a heavy fermion ground state at low temperatures.
\ybal~differs from its polymorph, \aybal, firstly in that superconductivity has only been found in \ybal~(below 80mK). Furthermore, measurements of the low temperature specific heat\cite{Macaluso1} show a larger linear specific heat coefficient saturates at above
150 mol$^{-1}$ K$^{-1}$ with the application of small magnetic fields\cite{NakatsujiYb1} in \ybal~compared to approximately 130 mol$^{-1}$ K$^{-1}$ \aybal\cite{Macaluso1}. This may point to stronger correlation effects in \ybal~that contribute to its superconductivity. Electron spin resonance measurements\cite{HolandaYb1} also suggest that \ybal~possesses a unique signature of both local moment and conduction electron behavior that has been attributed to its quantum criticality and is not observed in \aybal. Also, while \ybal~shows features of quantum criticality and non-Fermi liquid behavior at low temperatures and ambient pressure, \aybal~behaves as a conventional Fermi liquid\cite{TomitaYb1}.

In this investigation we consider the electronic and magnetic properties that may drive the intriguing behavior of both \ybal~and its polymorph \aybal. We utilize density functional methods with DFT+U to model the system. Importantly, these calculations differ from the itinerant $f$ electron calculation favored in a recent quantum oscillation study of the low temperature Fermi surface\cite{O'FarrellYb1}. However, a DFT+U approach allows us to consider the driving interactions behind the behavior of the system, in particular in a Kondo physics framework.
We also consider the relationship between these materials and related 4$f^{13}$ materials such as \yrs~and YbB$_{2}$. In particular we consider the coupling of the 4\textit{f} moments to the conduction sea and the proximity of \ybal~to magnetic instabilities.
%
\\

\begin{center}
\textbf{II. Crystal Structure}
\end{center}

\begin{figure}
\centering
\begin{tabular}{c}
\includegraphics[width=0.87\linewidth,angle=0]{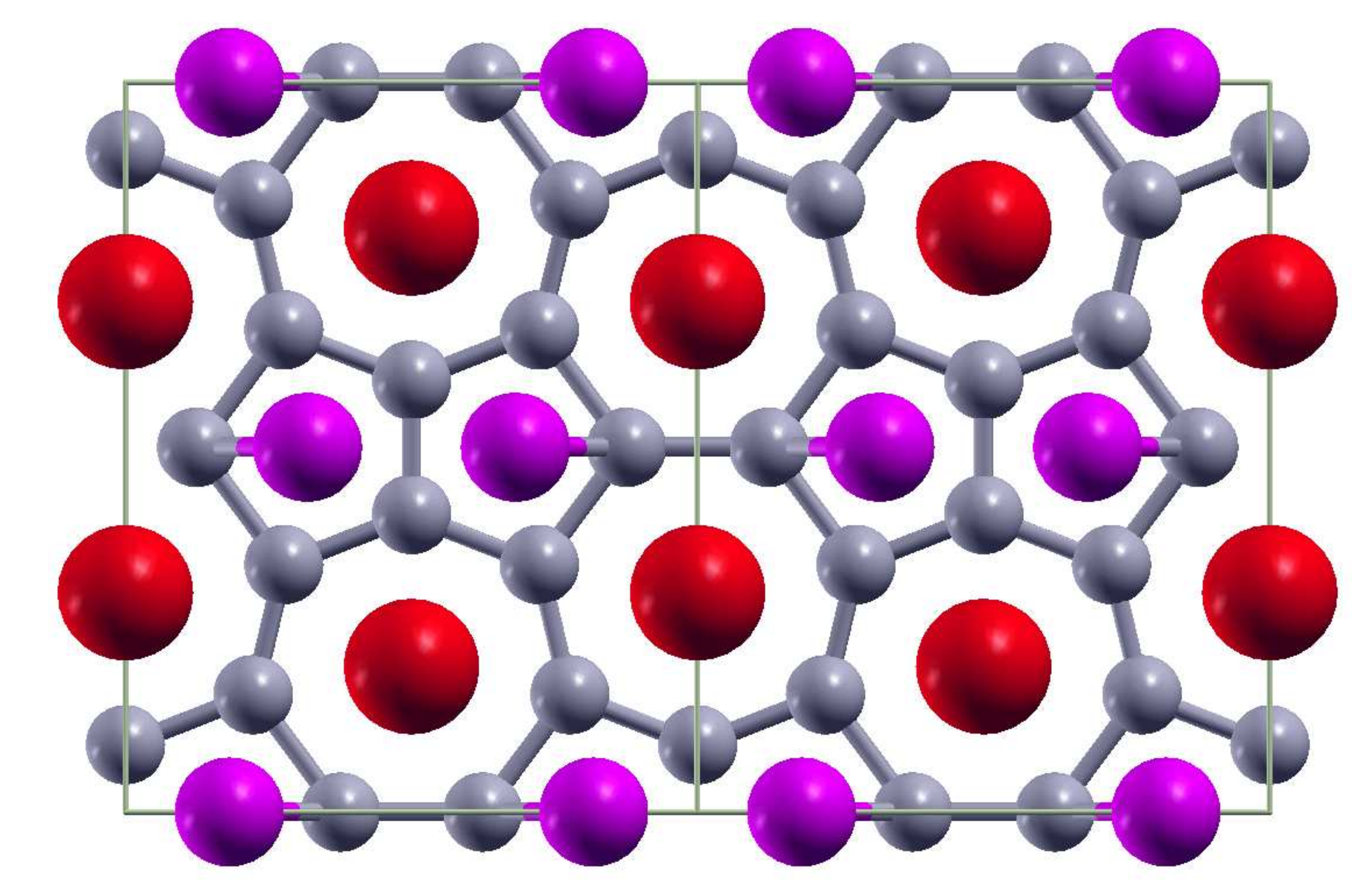} \\
\includegraphics[width=0.7\linewidth,angle=0]{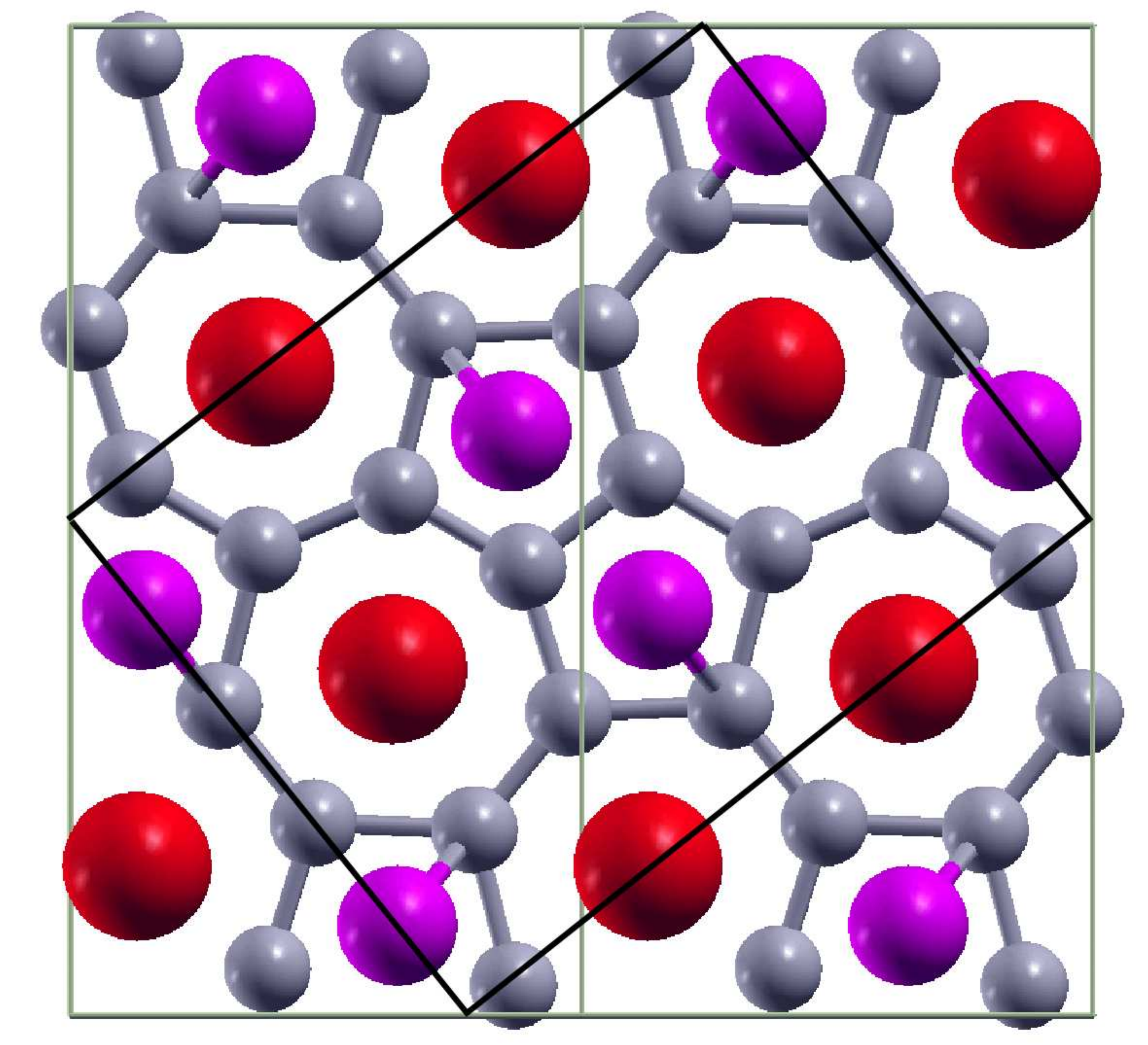} 
\end{tabular}
\caption{\label{fig:DNbFe2} (Color online) Crystal structure of (top) \textit{Cmmm} \ybal~and (bottom) \textit{Pbam} \aybal~viewed along the \textbf{c}-axis The green rectangular outlines show the conventional unit cells and we have doubled the unit cell along the \textbf{a}-axis. The black rectangle in the lower figure shows an outline of a unit cell in \aybal~that may be most easily compared to the conventional cell of the upper \ybal. }
\end{figure}

In this investigation we use the crystal structure extracted experimentally by Macaluso {\it et al.}\cite{Macaluso1} 
for both \ybal~and \aybal.
\ybal~crystallizes in the ThMoB$_{4}$ structure with space group \textit{Cmmm} and
\aybal~crystallizes in the YCrB$_{4}$ structure with space group \textit{Pbam}. Both space groups are orthorhombic. The lattice parameters are shown in Table~\ref{tab:cellParams}. We also show the crystal structures of both polymorphs in Fig.~\ref{fig:DNbFe2}. We have doubled the conventional cells of each structure along the \textbf{a}-axis to aid visualization of the bonding networks. The lower symmetry structure of \aybal~in the lower panel may be most easily compared to \ybal~via the construction of the approximate unit cell outlined in the lower panel of Fig.~\ref{fig:DNbFe2}. In comparing this new rotated cell in \aybal~with the structure of \ybal, the bonding networks become similar. However, subtle differences in symmetry mean that the transport, heat capacity, and the bonding in the two systems are different.

\begin{table*}
\caption{\label{tab:cellParams} Space group, cell parameters, atomic positions and site symmetries for $\alpha$-, $\beta$- and $\gamma$- YbAlB$_4$ as well as \yb.}
\begin{ruledtabular}
\begin{tabular}{ |c|c|c|c||c|c|}
  &  & $\alpha$ & $\beta$ & $\gamma$ & \yb  \\
\hline
Space group & & \textit{Pnam} (No.55) & \textit{Cmmm} (No.65) & ~~~~~~\textit{Pmmm} (No. 47)~~~~~~ & ~~~~~~\textit{P6/mmm} (No. 191)~~~~~~ \\
\hline
Lattice & a & 5.9220 & 7.3080 & 3.1346 & 3.2561 \\
constants & b & 11.4370 & 9.3150 & 5.4293 &  \\
(\AA) & c & 3.5060 & 3.4980 & 3.4980 & 3.7351 \\
\hline
~~~~~Volume (\AA$^{3}$/f.u.)~~~~~ &  & 59.55 & 59.53 & 59.53 & 34.29 \\
\hline
Atomic & Yb & 4\textit{g} (\textit{..m}) & 4\textit{i} (\textit{m2m}) & 1\textit{a} (\textit{mmm}) & 1\textit{a} (\textit{6/mmm}) \\
positions &  & (0.1294, 0.10543, 0) & (0, 0.30059, 0) & (0, 0, 0) & (0, 0, 0) \\ \cline{2-6}
(Wyckoff& Al & 4\textit{g} (\textit{..m}) & 4\textit{g} (\textit{2mm}) & 1\textit{f} (\textit{mmm}) &  \\
position,&  & (0.1387, 0.4096, 0) & (0.1816, 0, 0) & (1/2, 1/2, 0) &  \\ \cline{2-6}
site    & B1 & 4\textit{h} (\textit{..m}) & 4\textit{h} (\textit{2mm}) & 2\textit{p} (\textit{m2m}) & 2\textit{d} (\textit{-6m2}) \\
symmetry,&  & (0.2893, 0.3126, 1/2) & (0.1240, 1/2, 1/2) & (1/2, 1/6, 1/2) & (1/3, 2/3, 1/2) \\ \cline{2-6}
coordinates& B2 & 4\textit{h} (\textit{..m}) & 8\textit{q} (\textit{..m}) & 2\textit{n} (\textit{m2m}) &  \\
(x,y,z))&  & ~~~~~~(0.2893, 0.3126, 1/2)~~~~~~ & ~~~~~~(0.2232, 0.1609, 1/2)~~~~~~ & (0, 1/3, 1/2) &  \\ \cline{2-6}
& B3 & 4\textit{h} (\textit{..m}) & 4\textit{j} (\textit{m2m}) &  &  \\
&  & (0.3840, 0.0468, 1/2) & (0, 0.0920, 1/2) &  &  \\ \cline{2-6}
& B4 & 4\textit{h} (\textit{..m}) &  &  &  \\
&  & (0.4740, 0.1943, 1/2) &  &  &  \\
\end{tabular}
\end{ruledtabular}
\end{table*}

Both structures are orthorhombic with relatively short \textbf{c}-axis lattice parameters which will bear importance for our electronic structure results. Both structures also contain a layer of interconnected 5-member and 7-member boron rings intercalated with Al and Yb atoms. Each Yb atom is coordinated above and below by 7-member boron rings in both structures. These seven member rings dominate the crystal field environment of the Yb atom and therefore impact its orbital state. 

Crystallographically, the two polymorphs are primarily distinguished by the manner in which the 5-member and 7-member rings of the boron 
plane are interconnected. Both the 5-member and 7-member rings are irregular polygons. This irregularity is far more 
pronounced in the $\alpha$ structure. Therefore the intercalant Yb and Al atoms do not reside equally close to all 
boron atoms in the nearest boron ring in the two structures. For example, in Fig.~\ref{fig:DNbFe2} the bonds from 
the Al show the link to its nearest boron atom. In \ybal~these bonds lie along a single axis while there are two 
different orientations in $\alpha$-YbAlB$_4$.  Although this difference is unlikely to influence the orbital state 
of an individual Yb Atom, it may be influential upon the magnetic coupling between the Yb 4$f$ magnetic 
moments. The alteration in symmetry is then likely to influence the magnetic state of the system and in this way be 
important to understanding the properties of these materials. 


In this study we also consider the properties of two related crystal structures that both have 6-member coordination of the Yb ion. The first is a hypothetical structure, which we refer to as \gybal, and has the same c-axis lattice parameter and cell volume as \ybal, but a honeycomb boron sublattice that leads to a regular 6-member coordination of the Yb and Al ions. The second structure we compare to is \yb(YbYbB$_4$) which exists in the same hexagonal symmetry as MgB$_2$ (\textit{P6/mmm})\cite{Avila1}. The cell parameters for both of these structures are also shown in Table~\ref{tab:cellParams}. Experimentally in \yb~the Yb $f^{13}$ orbitals exhibit a Curie-Weiss like susceptibility at high temperatures and the compound is thought to order antiferromagnetically\cite{Avila1} at $T_N=5.6 \pm 0.2K$.

\begin{center}
\textbf{III. Electronic Structure Methods}
\end{center}
The electronic structures determined in this investigation are based on the all-electron approach to density functional theory using full-potentials in WIEN2k\cite{Blaha1}. The Brillouin zone integration was achieved
by the tetrahedron method typically using a $18\times18\times30$ \textbf{k}-point mesh. The radii of the muffin tins were set at 1.6$a_0$ for B, 2.11$a_0$ for Al and 2.15$a_0$ for Yb. We utilised $RK_{max}=8.0$.

We have calculated the electronic structure using the GGA-PBE\cite{PerdewYb1} correlation functional and with +U on the Yb sites for the 4$f$ orbitals. We employ $U=8$ eV and $J=1$ eV and spin-orbit coupling in a second variational approach. The `around mean field' double counting correction in WIEN2k was used.

\begin{figure}
\psfrag{ybalFrag}[0][0][2]{\textbf{\large \ybal~DOS}}
\psfrag{aybalFrag}[0][0][2]{\textbf{\large \aybal~DOS}}
\psfrag{ybFrag}[0][0][2]{\textbf{\large \yb~DOS}}
\psfrag{ybfFrag}[0][0][2.25]{\textbf{\large Yb-$f$}}
\psfrag{JzFrag}[0][0][2]{\textbf{\large Yb-$f(J_z=-5/2)$}}
\centering
\begin{tabular}{c}
\includegraphics[width=1\linewidth,angle=0]{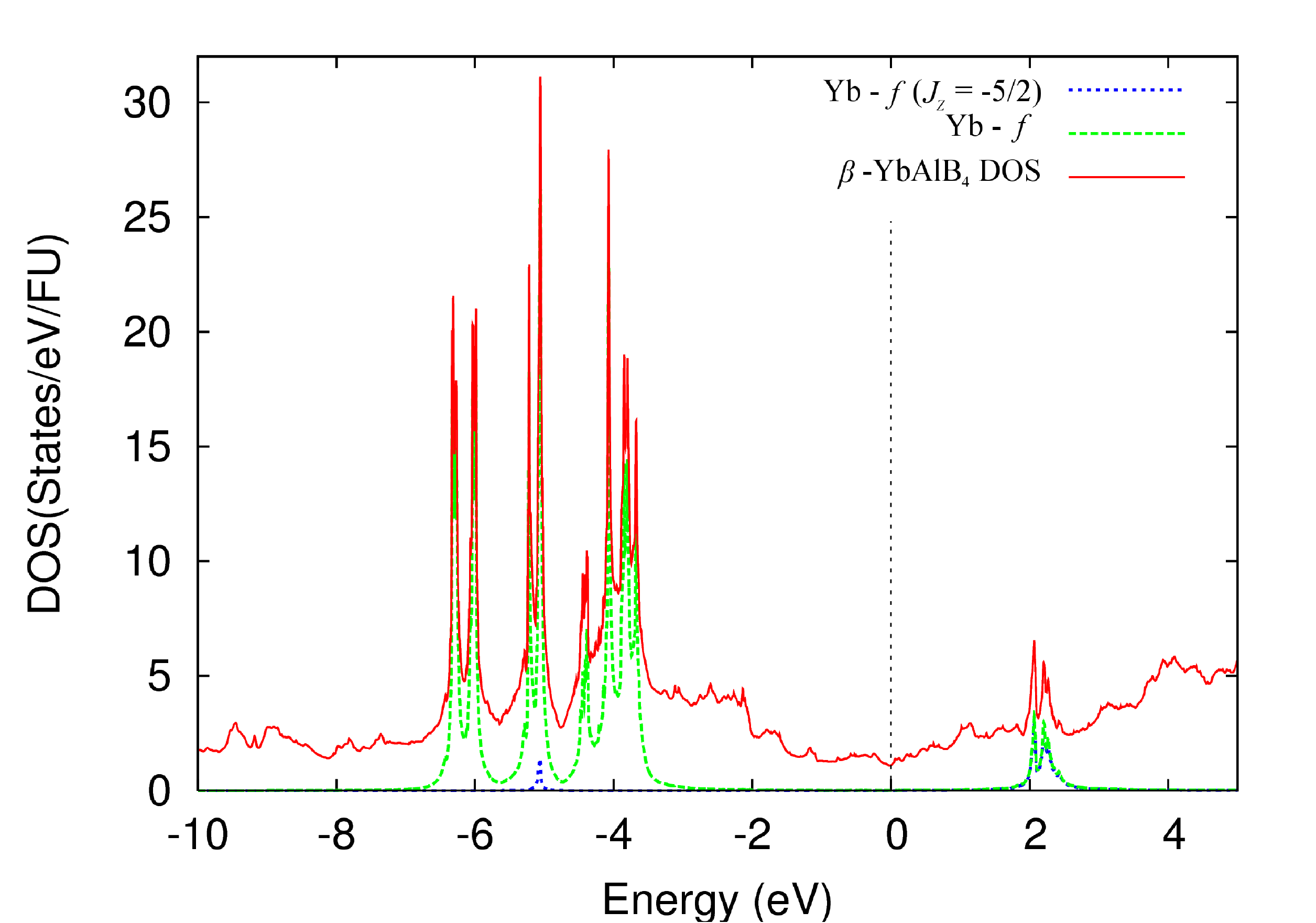} \\
\includegraphics[width=0.97\linewidth,angle=0]{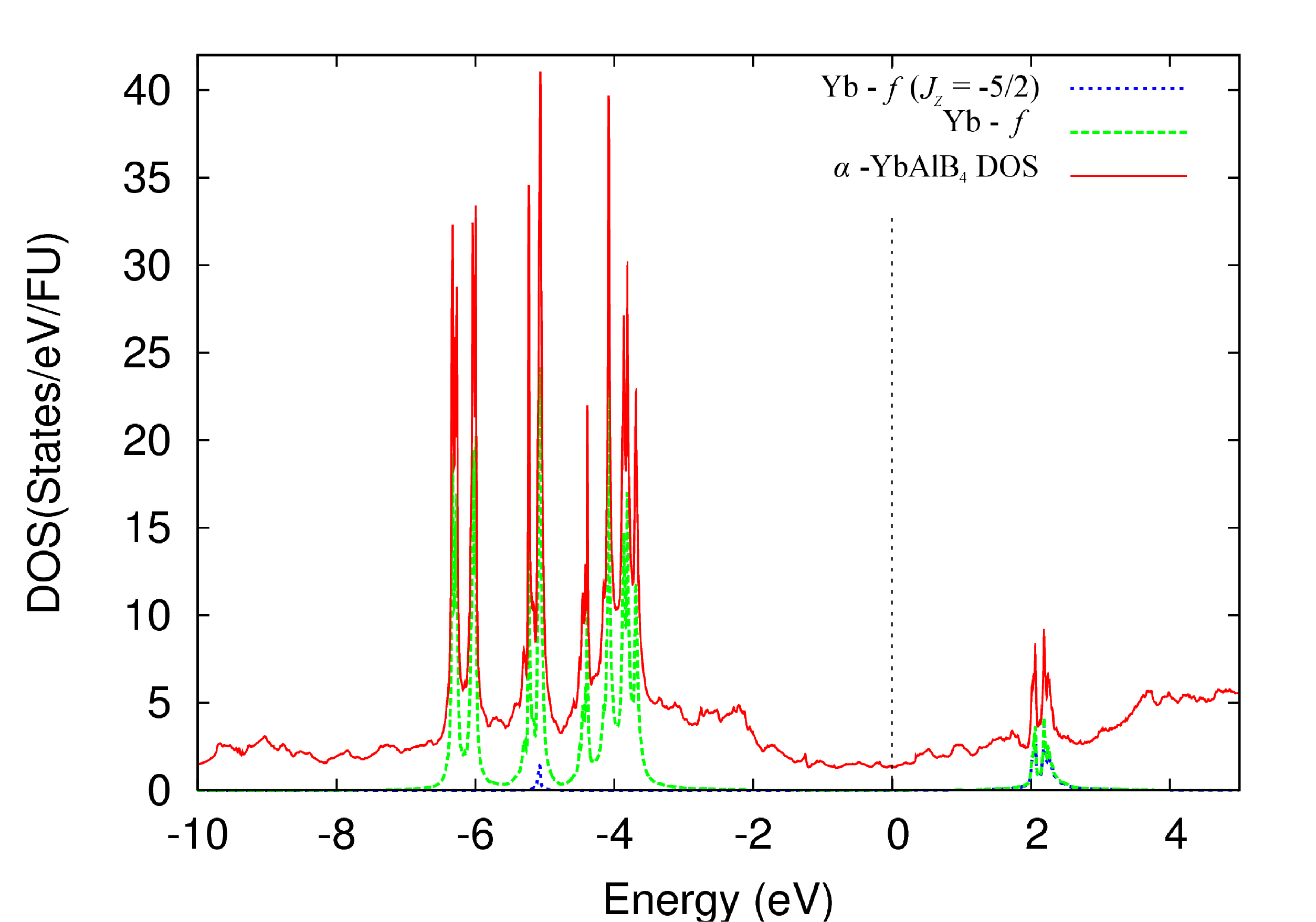} \\
\end{tabular}
\caption{\label{fig:ybalDOS} (Color online) Density of states for \ybal~(top) and \aybal~(bottom).}
\end{figure}

In this section we compare and contrast the electronic structure of \ybal~and \aybal~with \gybal~and \yb. We begin by showing the calculated density of states for \ybal~and \aybal~in Fig.~\ref{fig:ybalDOS}. In all cases the large peaks in the DOS are dominated by the Yb 4$f$ contributions. The overall structure of the DOS for \ybal~and \aybal~are similar. For both, a large manifold of states due to the filled \textit{f} orbitals lies between -7 eV and -3 eV below the Fermi level.  The states between -2 eV and +2 eV about the Fermi level are dominated by itinerant states that have little charge within the muffin tins. Visualization of the charge density shows that most of the itinerant charge is associated with the boron layers. The empty 4$f^{13}$ hole state sits just beyond 2 eV above the Fermi level. It is possible to obtain GGA+U solutions with a variety of different hole states (however, not all can be achieved). We find that the $J_z=-5/2$ hole orbital always leads to the lower energy. 

The $J_z=-5/2$ hole is consistent with the result predicted by an approximate treatment of the crystal field symmetry of the Yb site\cite{Nevidomskyy1}. A low lying doublet of $m_J= \pm 5/2$ was predicted and shown to give a good fit to the measured magnetic susceptibility at high temperatures. Furthermore, our total energy calculations suggest that the energy scale for non-Ising flops from this state is at least 0.3eV ($\sim$ 3481K) which implies that thermal flops in this system are likely to be Ising-like. Calculating the expectation value of the Ising-like moment gives 3.4$\mu_B$ which is comparable to the measured values of the effective moment of 2.9$\mu_B$\cite{Macaluso1} and 3.1$\mu_B$\cite{NakatsujiYb1}. In our calculations the order is ferromagnetic and the magnetization is directed along the c-axis which is the easy axis in the measured magnetization\cite{Macaluso1}. We note that the calculated specific heat coefficients of 1.21 and 1.54 mJ/mol K$^2$ for \ybal~and \aybal~correspond to large mass enhancements $\sim$100. This implies the presence of a low temperature heavy fermion ground state that is not represented by the bare DFT+U calculation.

\begin{figure}
\psfrag{ybalFrag}[0][0][2]{\textbf{\large \ybal~DOS}}
\psfrag{aybalFrag}[0][0][2]{\textbf{\large \aybal~DOS}}
\psfrag{gybalFrag}[0][0][2]{\textbf{\large \gybal~DOS}}
\psfrag{ybFrag}[0][0][2]{\textbf{\large \yb~DOS}}
\psfrag{ybfFrag}[0][0][2.25]{\large \textbf{Yb-$f$}}
\psfrag{JzFrag}[0][0][2]{\large \textbf{Yb-$f(J_z=-5/2)$}}
\centering
\begin{tabular}{c}
\includegraphics[width=1\linewidth,angle=0]{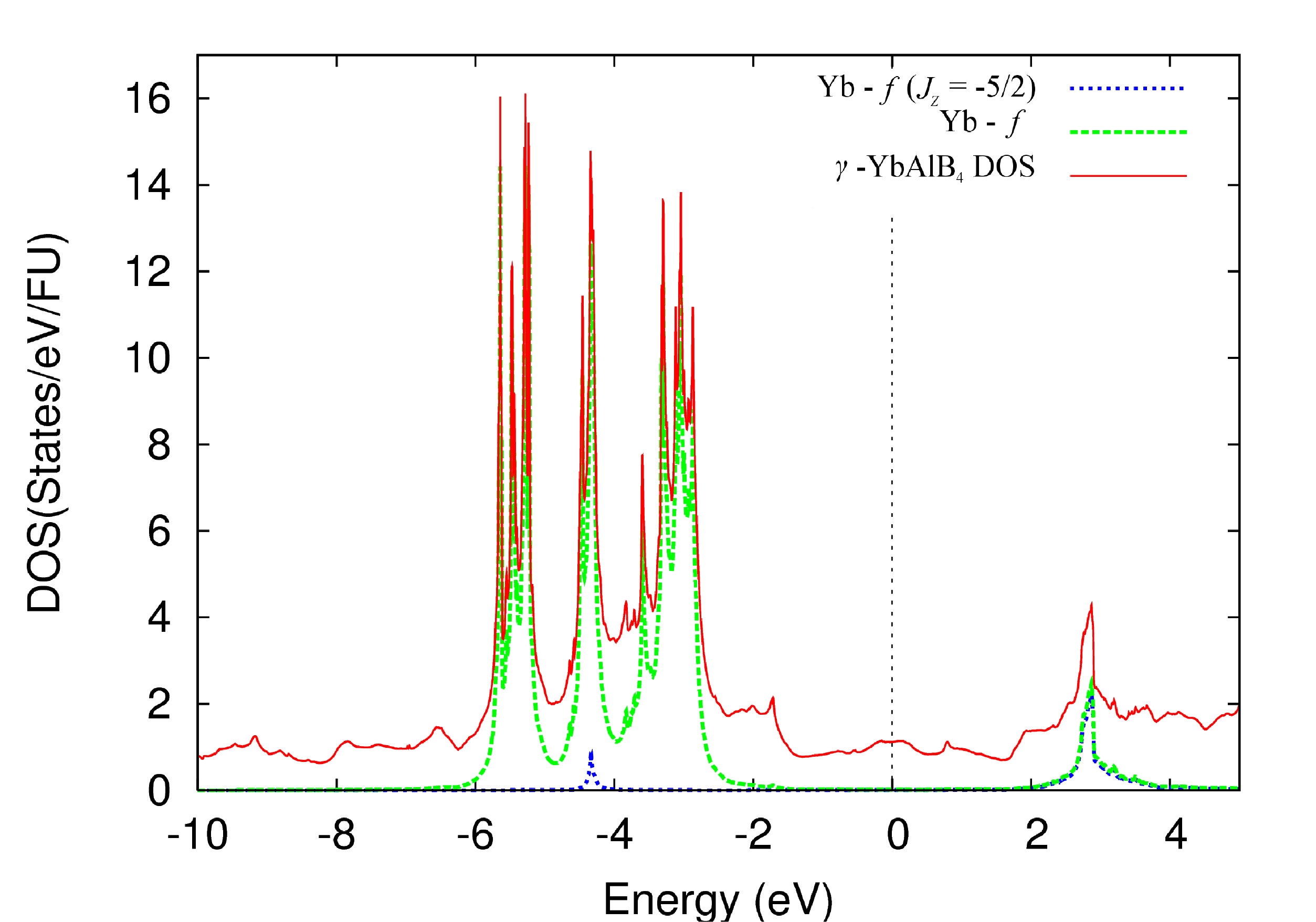} \\
\includegraphics[width=1\linewidth,angle=0]{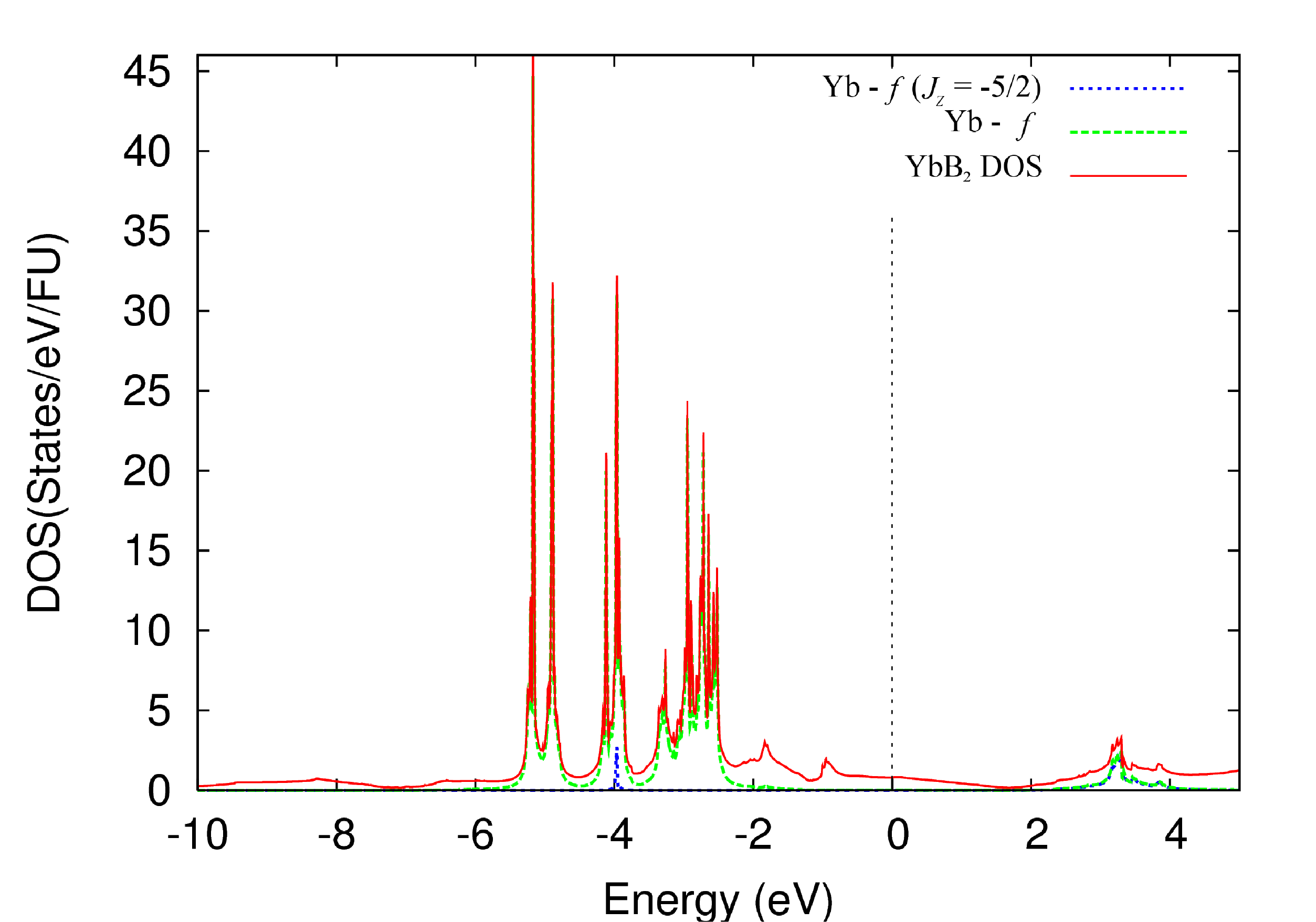} \\
\end{tabular}
\caption{\label{fig:gybalYbB2DOS} (Color online) Density of states for \gybal~(top) and \yb (bottom).}
\end{figure}

In Fig.~\ref{fig:gybalYbB2DOS} we show the DOS for \gybal~and \yb. Both exhibit a similar localised 4$f^{13}$ hole structure, but the hole state lies significantly further from the Fermi level than for both polymorphs of YbAlB$_4$ at approximately +3 eV. We find the entire manifold  of \textit{f}-states, both above and below the Fermi level, is shifted upwards and this is a result of the differing crystal field produced by the coordination of 6-member boron rings as well as the effective volume about the Yb site. Interestingly \yb~seems to exhibit a very nearly gapped density of states at $\approx+1.75$ eV. Together these indicate that the 4$f$ hole may be less strongly linked by hybridization and spin-orbit effects with the itinerant states near the Fermi level.

\begin{center}
\textbf{IV. Competing States and Quantum Criticality}
\end{center}

We digress from the electronic structure to discuss briefly some analysis relating to heavy fermion physics.
In the Doniach picture of heavy fermion systems two scales may be considered to dominate the physics: the Kondo scale $T_{K}$ of local moment screening and the scale for the 4$f$ moments to magnetically order such as $T_{RKKY}$. By multiplying by the density of states at the Fermi level these may be expressed in dimensionless form as:
\begin{equation}
\rho(E_F) T_{K} \propto \textrm{exp} \left(- \frac{E_0}{\Delta}  \right) 
\end{equation}

\begin{equation}
\rho(E_F) T_{RKKY} \propto  (\frac{\Delta}{E_0})^2 \\
\end{equation}

\begin{equation}
 \text{where~~~~} J \rho(E_F) = \frac{\Delta}{E_0}
\end{equation}
\noindent Here $\Delta$ and $E_0$ are respectively the width and position with respect to the Fermi level $E_F$ of the 4$f$ level. Here $J$ is the local moment - itinerant electron coupling as in the Kondo model. We note that this Doniach approximation does not account for the degeneracies of the 4$f$ state in Yb\cite{ReadYb1}. The type of magnetic order and dimensionality may have further influences on the system\cite{IrkhinYb1, IrkhinYb2, IrkhinYb3}. However, within the model we may gain insight into the drivers of the Kondo physics in these systems.

Theoretical\cite{Nevidomskyy1} and experimental work using both doping and pressure tuning\cite{TomitaYb1} indicate that \ybal~is proximate to an antiferromagnetic instability. Experimentally, stoichiometric \yb~is thought to order antiferromagnetically\cite{Avila1} at $T_N=5.6 \pm 0.2K$. Given the similar crystal structures of \yb~and \ybal, we may compare these two compounds to determine which characteristics may promote the antiferromagnetism in \yb~and the quantum criticality of \ybal.

In order to tune \yb~away from antiferromagnetism towards a quantum critical point we need to effect an enhancement of $T_K$ over $T_{RKKY}$. Due to the exponential dependence of $T_K$ this may be achieved by an increase in the quantity $J \rho(E_F) = \frac{\Delta}{E_0}$. We may, for example, tune this quantity by the application of pressure, $p$. The application of pressure in Yb systems\cite{BoursierYb1, YlvisakerYb1, GoltsevYb1} will increase $\Delta$ due to increased hybridization and also increases $E_{0}$ since the valence fluctuation is between Yb$^{2+}$ and Yb$^{3+}$. Essentially, this is because the fluctuation involves the addition of an $f$ electron and therefore the effective energy barrier to the process, $E_{0}$, is smaller when the volume about the Yb ion is larger. Experimentally, the behavior of $E_{0}$ dominates\cite{MederleYb1, PlesselYb1} and we are tuned deeper into the antiferromagnetic state with increasing $p$. Therefore, we typically require an effective negative pressure, or volume increase, in order to tune from a magnetic state towards a quantum critical point. Experimentally, an effective negative pressure has so far not been achieved by chemical pressure in these systems, but we may simulate the effect by increasing the unit cell volume in our electronic structure calculations.

\begin{figure}
\psfrag{ybalFrag}[0][0][2]{\textbf{\large \ybal~DOS}}
\psfrag{aybalFrag}[0][0][2]{\textbf{\large \aybal~DOS}}
\psfrag{ybFrag}[0][0][2]{\textbf{\large Total DOS}}
\psfrag{ybfFrag}[0][0][2.25]{\textbf{\large Yb-$f$}}
\psfrag{JzFrag}[0][0][2]{\textbf{\large Yb-$f(J_z=-5/2)$}}
\centering
\begin{tabular}{c}
\includegraphics[width=1\linewidth,angle=0]{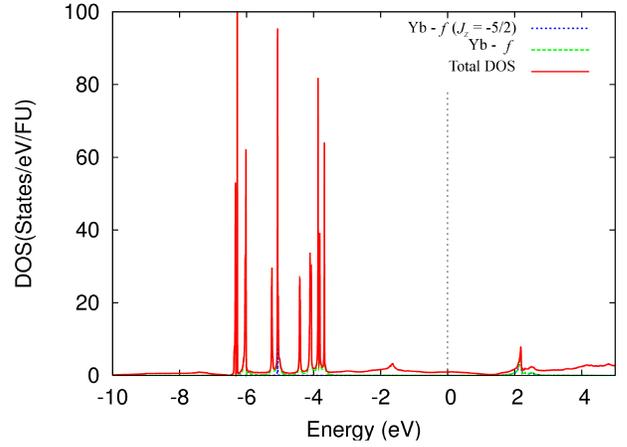} \\
\end{tabular}
\caption{\label{fig:VolDepYbB2DOS} (Color online) Density of states for \yb~at the expanded unit cell volume: 46.68\AA$^{3}$.}
\end{figure}

In Fig.~\ref{fig:VolDepYbB2DOS} we show the density of states for the structure of \yb~at an expanded unit cell volume of 46.68\AA$^{3}$. The volume about the Yb ion is increased. As a result $E_{0}$ falls and $\Delta$ also decreases compared to \yb~ at its experimental volume in Fig.~\ref{fig:gybalYbB2DOS}. In this structure the value of $E_{0}$ is approximately the same as that in \ybal. However, the width, $\Delta$, in the expanded \yb~structure is much smaller than that found in \ybal. As a result with such a volume tuning, $\Delta$ is likely to be small by the time we tune to a critical point.

In Table~\ref{tab:kondoParams} we compare the value of $\frac{\Delta}{E_0}$ for several important structures. An important feature from the table is the ability of \ybal~and \aybal~to maintain a large hybridization, $\Delta$, even with a larger Yb-B distance compared to \yb. This difference suggests that parameters beyond the volume about the Yb ion are important. The angle of inclination of the Yb ion to the boron ring may be pivotal to producing an enhanced $\frac{\Delta}{E_0}$ (and Kondo scale) while maintaining a large $\Delta$. 

\begin{table*}
\caption{\label{tab:kondoParams} The value of the parameter $\frac{\Delta}{E_0}$. To estimate $\Delta$ we have used the full width at half maximum of the 4\textit{f} hole peak in the density of states.}
\begin{ruledtabular}
\begin{tabular}{cccccc}
 Structure & Yb-B Distance (\AA) & Yb-B Angle & $\Delta$ & $E_0$ & $\frac{\Delta}{E_0}$  \\
\hline
\ybal & 2.70 & ~48$^{\circ}$ & 0.25 & 2.2 & 0.11 \\
\aybal & 2.73 & ~48$^{\circ}$ & 0.23 & 2.2 & 0.10 \\
$\gamma$-YbAlB$_4$ & 2.516 & ~46$^{\circ}$ & 0.2 & 2.8 & 0.07 \\
\yb & 2.65 & ~45$^{\circ}$ & 0.2 & 3.2 & 0.06 \\
\end{tabular}
\end{ruledtabular}
\end{table*}

\begin{figure}
\includegraphics[ width=0.5\textwidth]{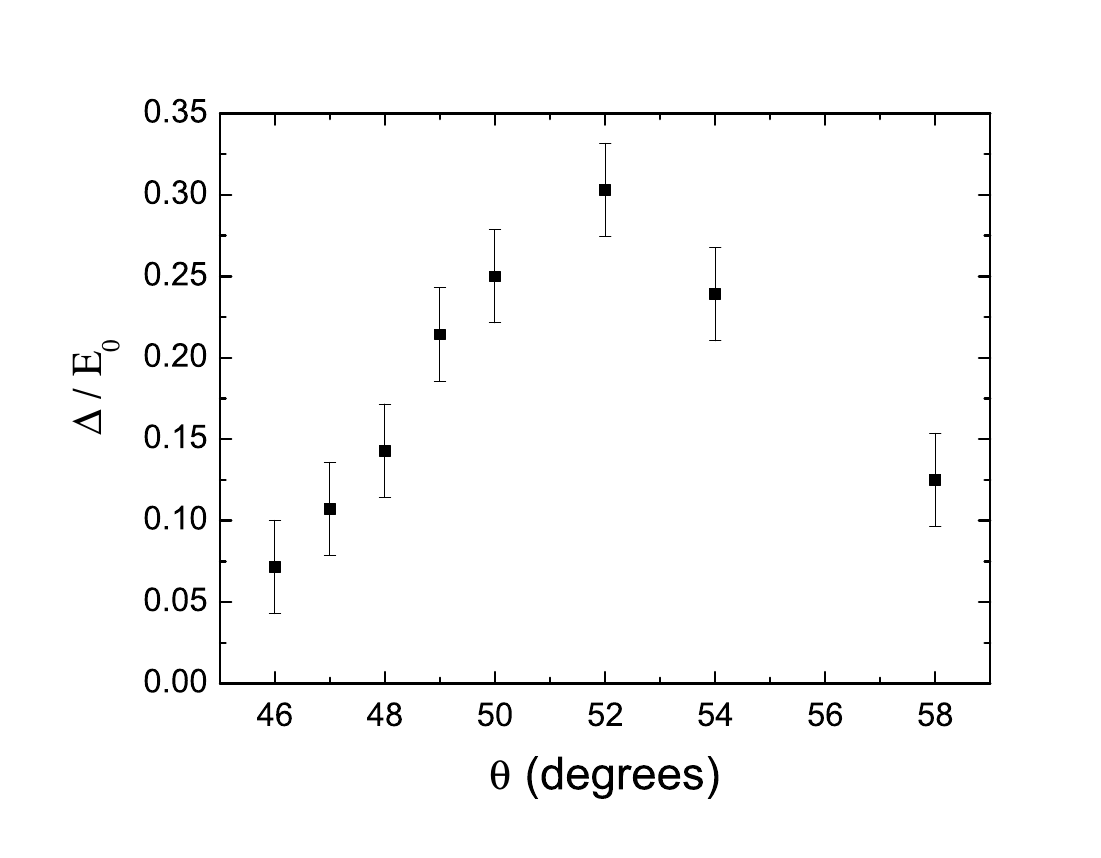}
\caption{\label{fig:GammaKondoVsAngle} (Color online) Plot of the quantity $\Delta / E_0$ versus the angle subtended by the Yb ion to the surrounding boron ring in $\gamma^*$-YbAlB$_4$. $\Delta / E_0$ is important to the determination of the Kondo scale and Table~\ref{tab:kondoParams} indicates that the structure of \ybal~lies nearest to the optimum.}
\end{figure}

To probe the importance of the angle, $\theta$, subtended by the Yb ion to the boron ring we utilize our hypothetical $\gamma$-YbAlB$_4$ structure. We do this by altering the aspect ratio of its unit cell to vary $\theta$, while retaining a constant unit cell volume to create a set of structures that we denote as $\gamma^*$-YbAlB$_4$. In Fig.~\ref{fig:GammaKondoVsAngle} we plot the value of $\Delta / E_0$ as a function of $\theta$. We see that the peak value is centered about 52$^{\circ}$, which is the angle subtended by the proposed ground state hole state $| m_J = \pm 5/2 > $. Importantly, while the angle $\theta$ is altered we find little variation in $E_0$ and it is the large changes in the hybridization, $\Delta$, that drive the strong dependence of $\Delta / E_0$ on angle. Therefore, \ybal~and \aybal~enhance their Kondo scale not only by increasing the effective Yb-B distance (to decrease $E_0$), but also approach the optimum angle for hybridization. It is the alternation of 5-member coordinated Al sites with 7-member coordinated Yb sites that allows \ybal~and \aybal~to form this advantageous angle of inclination. In this sense we may consider \ybal~and \aybal~to be \textit{symmetry tuned} Kondo enhanced materials. In \ybal~this Kondo enhancement drives the material to quantum criticality.

Interesting future work with a fuller treatment of many-body effects with dynamical mean field theory\cite{ShimYb1,MatsumotoYb1}, may attempt to demonstrate this relationship between crystal structure and the heavy fermion ground state as has been shown in elemental Yb under the influence of pressure\cite{YlvisakerYb1}. We also note that due to the nature of the DFT+U method, the absolute value of the parameter $\Delta / E_0$ may be sensitive to the value of $U$ used in the calculations. However, the trend in $\Delta / E_0$ as a function of the geometric change $\theta$ is consistent for different values of $U$.

In this framework, we may also discuss why \ybal~is the only discovered example of an Yb based heavy fermion superconductor, while many Ce based examples exist. We proceed by considering pressure tuning resulting in uniform volume contraction.
The effect of pressure on $\frac{\Delta}{E_0}$ will be different for Ce and Yb compounds. For Ce compounds the application of pressure decreases $E_{0}$, but increases $\Delta$. This is because the valence fluctuations are between Ce$^{3+}$ and Ce$^{4+}$. This fluctuation involves the removal of an $f$ electron and therefore the effective energy barrier, $E_{0}$, is small when the volume about the Yb ion is smaller. In contrast the application of pressure in Yb based compounds increases both $E_{0}$ and $\Delta$. In some sense these quantities are working against each other in terms of their Kondo physics in Yb compounds. For this reason at the quantum critical point, it is more likely that a narrow $\Delta$ will exist. Since $\Delta$ is indicative of the coupling to the itinerant electrons, then a lower tendency to superconductivity may be found in Yb based systems.

\begin{center}
\textbf{V. Band structure}
\end{center}
Critical to the operation of heavy fermion systems is the interaction of the localized \textit{f}-electrons with the conduction sea. This interaction leads to the dramatic mass enhancements and itinerant \textit{f}-electrons in these systems via the Kondo effect. The RKKY interaction may also give rise to a significant scale for magnetic coupling that drives their magnetic properties.

\begin{figure}
\includegraphics[ width=0.5\textwidth]{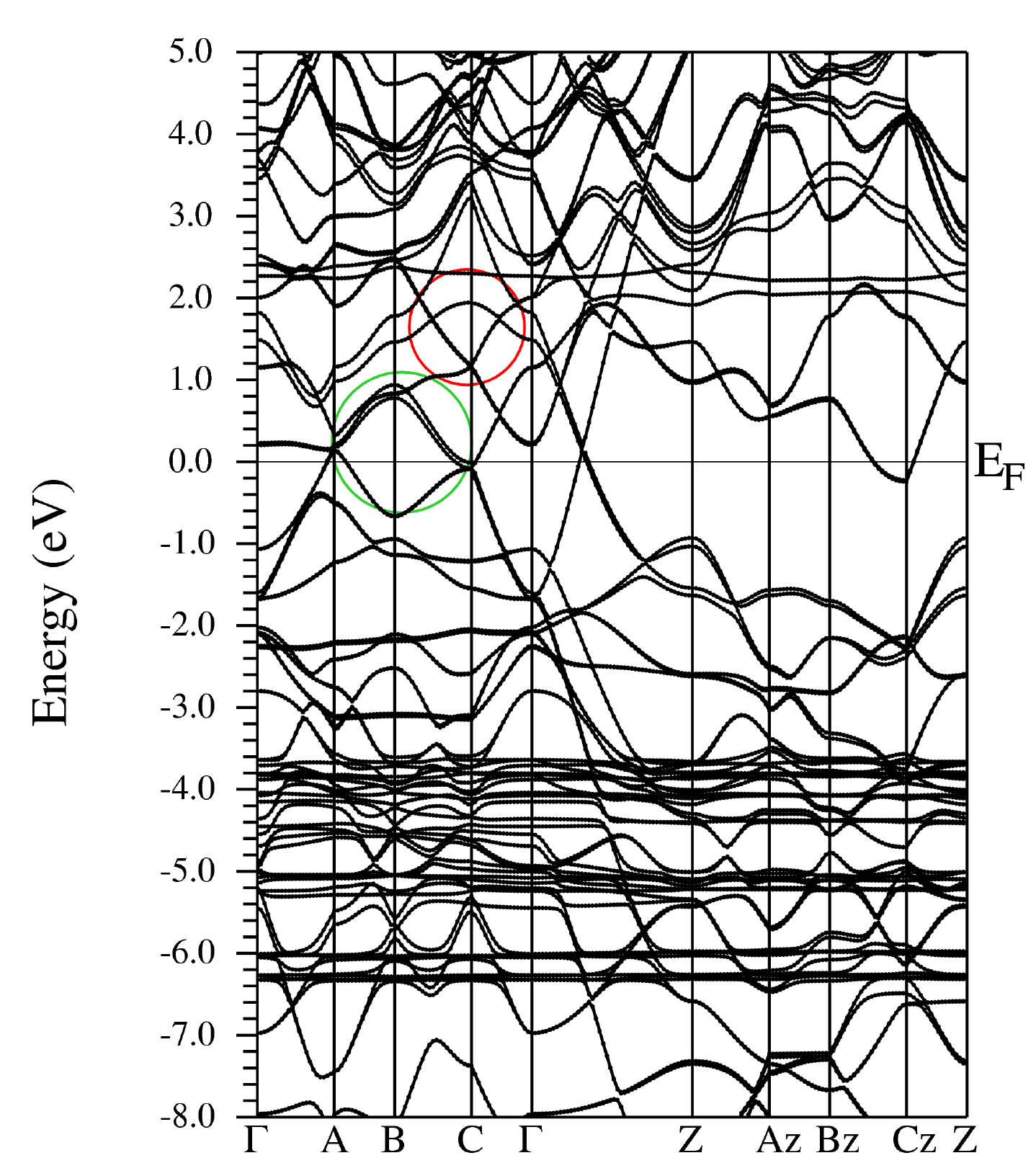}
\caption{\label{fig:bYbAlB4LDAPUBSConvCell} (Color online) Band structure for $\beta$-YbAlB4. The red circle indicates the location of a strongly hybridized \textit{f}-hole band. The green circle shows the location of strongly spin-split bands near the Fermi level (many splittings are not visible). Both indicate strong interaction of the \textit{f}-hole with the itinerant electrons.}
\end{figure}

In Fig.~\ref{fig:bYbAlB4LDAPUBSConvCell} we show the band structure for \ybal. We have elected to plot the dispersion along the symmetry directions for the reciprocal conventional cell. ($\mathbf{\Gamma}=(0~0~0)$, $\mathbf{A}=(1/2 ~0 ~0)$, $\mathbf{B}=(1/2 ~1/2 ~0)$, $\mathbf{C}=(0 ~1/2 ~0)$, $\mathbf{Z}=(0 ~0 ~1/2)$, $\mathbf{Az}=(1/2 ~0 ~1/2)$, $\mathbf{Bz}=(1/2 ~1/2 ~1/2)$ and $\mathbf{Cz}=(0 ~1/2 ~1/2)$, in Cartesian coordinates. For \ybal~there are two Yb atoms in the unit cell which lead to the appearance of two flat Yb bands at approximately +2.2 eV above the Fermi level. The remaining 4$f$ states are located in a manifold between -7 eV and -3 eV below the Fermi level. With SO coupling these 4$f$ bands are spin mixed and split into a 4$f_{5/2}$
complex and a 4$f_{7/2}$ complex separated by the spin-orbit
splitting of roughly 1.5 eV. These manifolds are further split by the anisotropy of the Coulomb interaction\cite{JohannesYb1}. This along with a Hund's rule splitting due to the unoccupied 4$f$-hole makes identifying the 4$f_{5/2}$ and 4$f_{7/2}$ states a challenging task.

Two important features of the band structure have been highlighted in Fig.~\ref{fig:bYbAlB4LDAPUBSConvCell}. The red circle, centered at 1.5 eV at point C, indicates the location of a strongly hybridized \textit{f}-hole band. In particular the archway shaped band shows strong \textit{f}-character in a partial charges analysis. This \textit{f}-character remains strong throughout its dispersion in the plane, and demonstrates a strong hybridization between the \textit{f}-hole and the conduction states. In contrast such hybridization is almost entirely absent in the related quantum critical compound \yrs\cite{WiggerYb1}. 

In this band structure of \ybal~the Yb moments are in a ferromagnetic configuration. This allows us to see the interaction of the Yb moment with the bands associated with itinerant electrons. The green circle, centered just above E$_F$ at point B, shows the location of strongly spin-split bands near the Fermi level. Both indicate strong interaction of the local \textit{f}-hole with the itinerant electrons. This has the benefit of providing a measure of the degree of
Kondo coupling of the Yb moment to the Fermi surfaces,
because the exchange splitting of these bands reflects
the coupling of the local moment to the itinerant bands. The exchange splitting of this band is approximately 100 meV, far larger than that found in \yrs \cite{WiggerYb1}. Consistent with experiment, the $T^*$, coherence temperature, in \ybal \cite{NakatsujiYb1} is $\approx200K$, almost an order of magnitude larger than that in \yrs\cite{TrovarelliYb1,CustersYb1}. This large Kondo coupling may be critical in the behavior of these materials and is consistent with large estimates of $T^*$ from the measured specific heat\cite{Macaluso1}. As can be seen, however, the band crossing the Fermi level at \textbf{Cz} possesses very little exchange splitting. Therefore the Kondo coupling is highly anisotropic and likely to vary around the Fermi surface. This is consistent with the varying mass enhancements inferred from recent quantum oscillation studies\cite{O'FarrellYb1}.

\begin{figure}
\includegraphics[ width=0.5\textwidth]{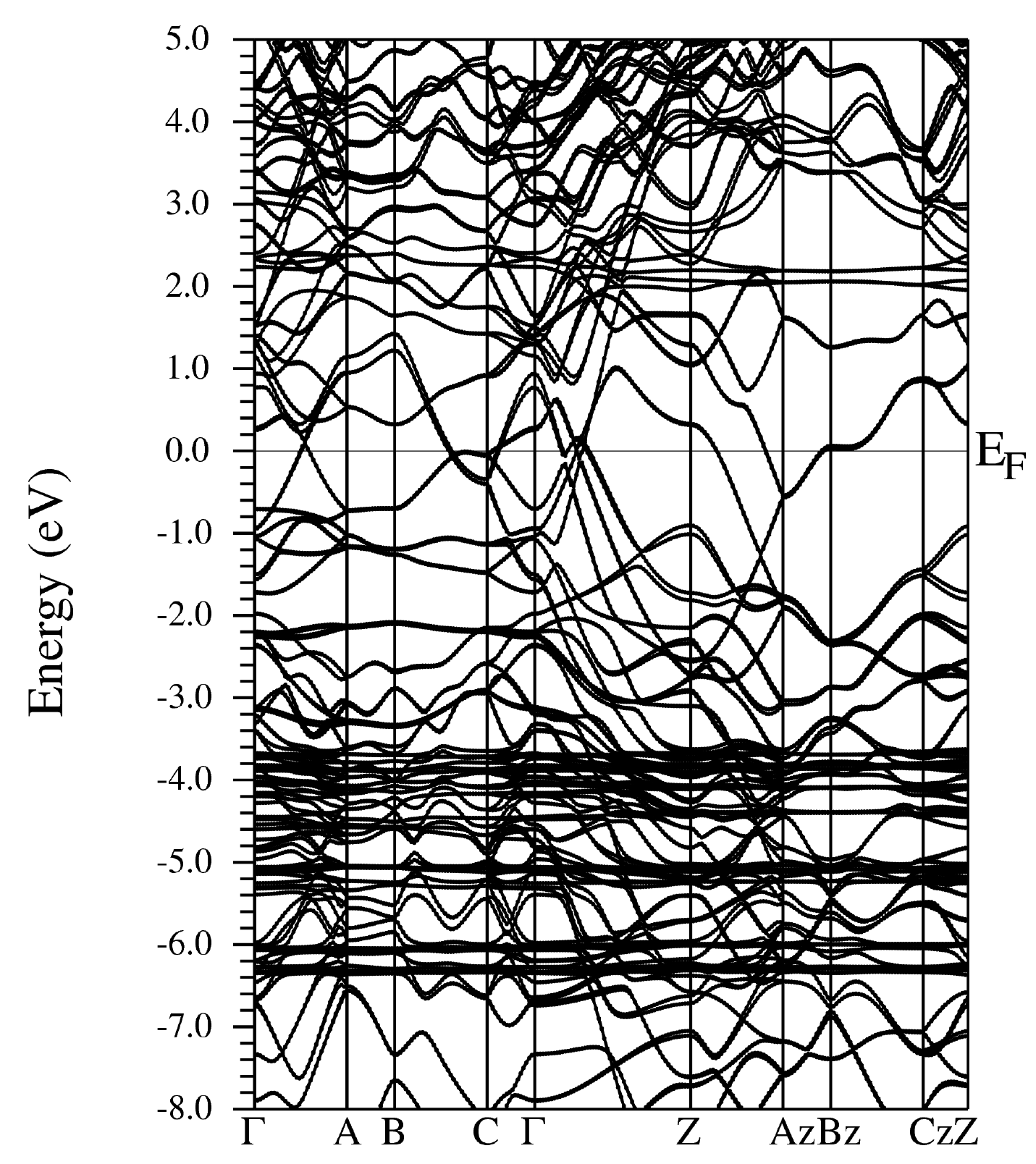}
\caption{\label{fig:aYbAlB4LDAPUBS} (Color online) Band structure for $\alpha$-YbAlB4.}
\end{figure}

In Fig.~\ref{fig:aYbAlB4LDAPUBS} we show the band structure for \aybal. Again the Yb moments are in a ferromagnetic configuration. We note that the band structure has twice as many bands as \ybal~due to the fact that its primitive unit cell contains 4 formula unit cells for \aybal, while for \ybal~there are only two. This leads to the appearance of increased complexity, yet the general structure of the two cases is similar. The position of the \textit{f}-hole states at +2.2 eV above the Fermi level and the location of the 4$f_{5/2}$ and 4$f_{7/2}$ multiplets between -6.5 eV and -3.5 eV below the Fermi level are similar. This similarity is due to the similar local symmetry of the Yb environment. The \aybal~band structure also shows evidence for spin split bands near the Fermi level. An example is provided by the spin split band at $\Gamma$ approximately +0.8 eV above the Fermi level.

In Fig.~\ref{fig:aYbAlB4LDAPUBS} we show the band structure for \yb. In this case there is only one formula unit per unit cell. Therefore there are fewer bands in the band structure and the Yb moment is necessarily in a ferromagnetic configuration. The band structure is highly dispersive in all directions. As expected from the density of states the \textit{f}-hole level lies further from the Fermi level than in the cases of \ybal~and \aybal. Also, there is less evidence for the presence of spin split itinerant bands near the Fermi level. These features are indicative of a smaller Kondo scale in \yb.
\begin{figure}
\includegraphics[ width=0.5\textwidth]{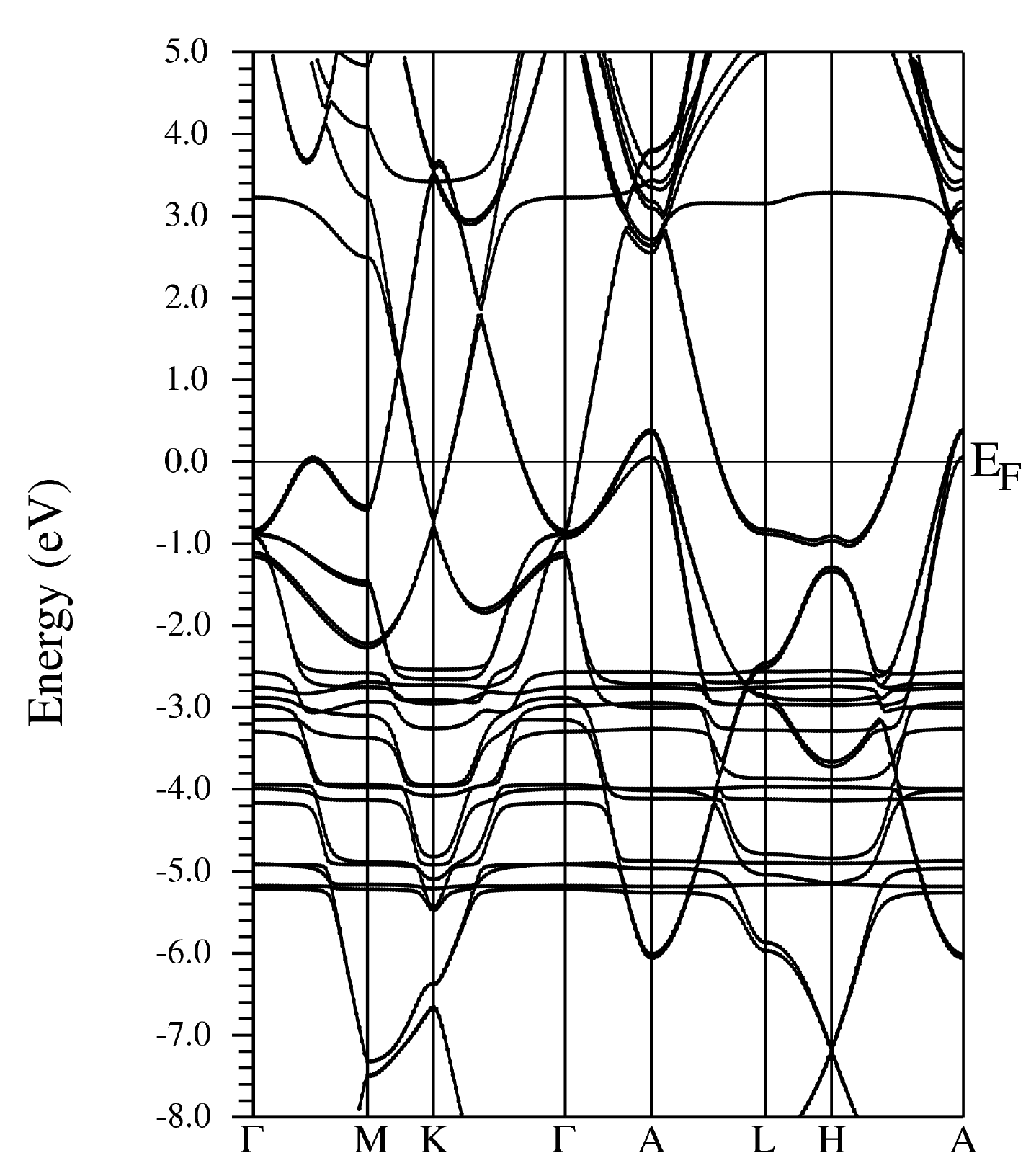}
\caption{\label{fig:YbB2LDAPUBS} (Color online) Band structure for \yb~in the hexagonal Brillouin zone.  }
\end{figure}

\begin{figure}
\centering
\begin{tabular}{c}
(a) \ybal \\
\includegraphics[width=1\linewidth,angle=0]{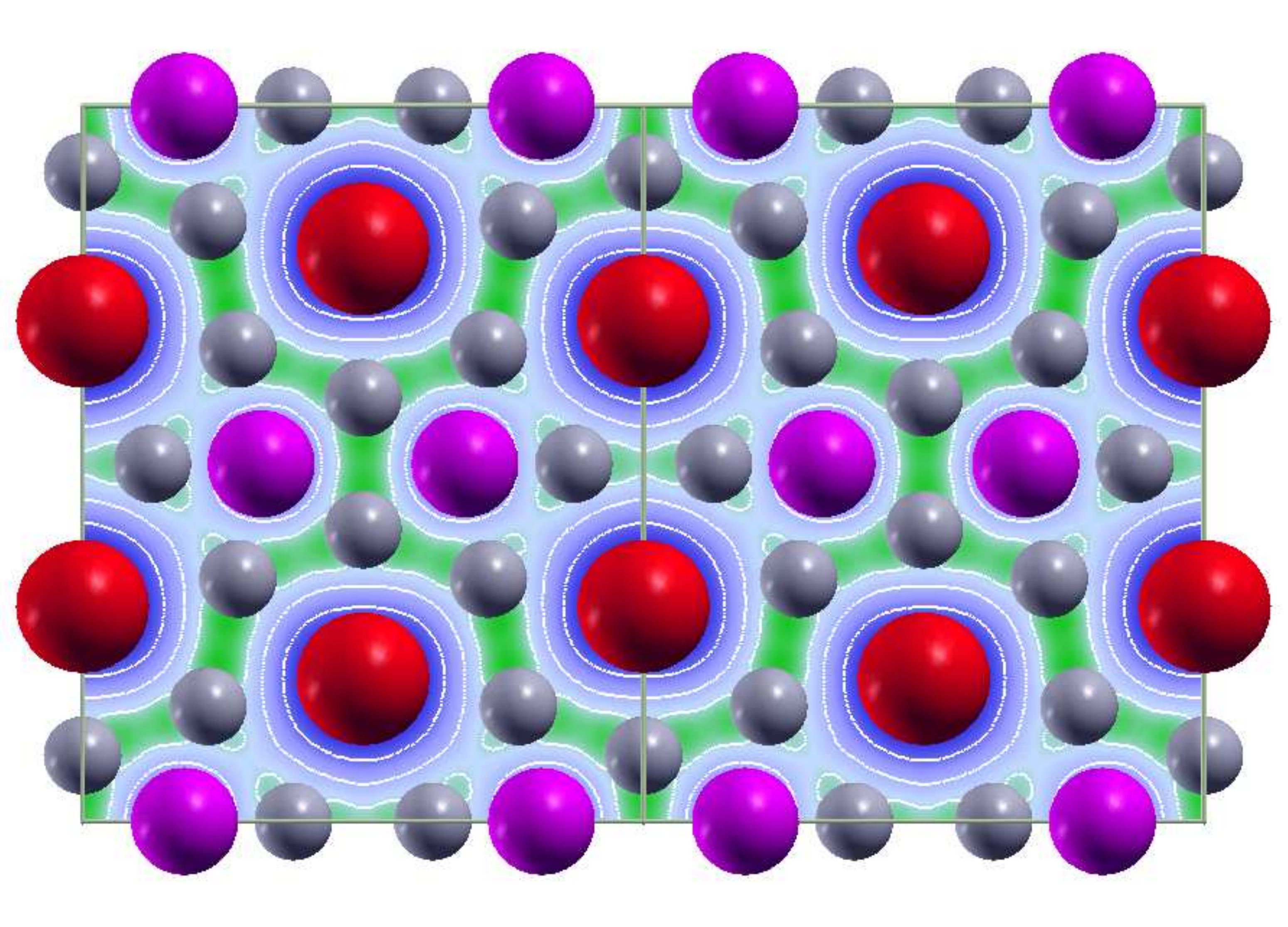}  \\
\\
(b) \aybal \\
\includegraphics[width=1\linewidth,angle=0]{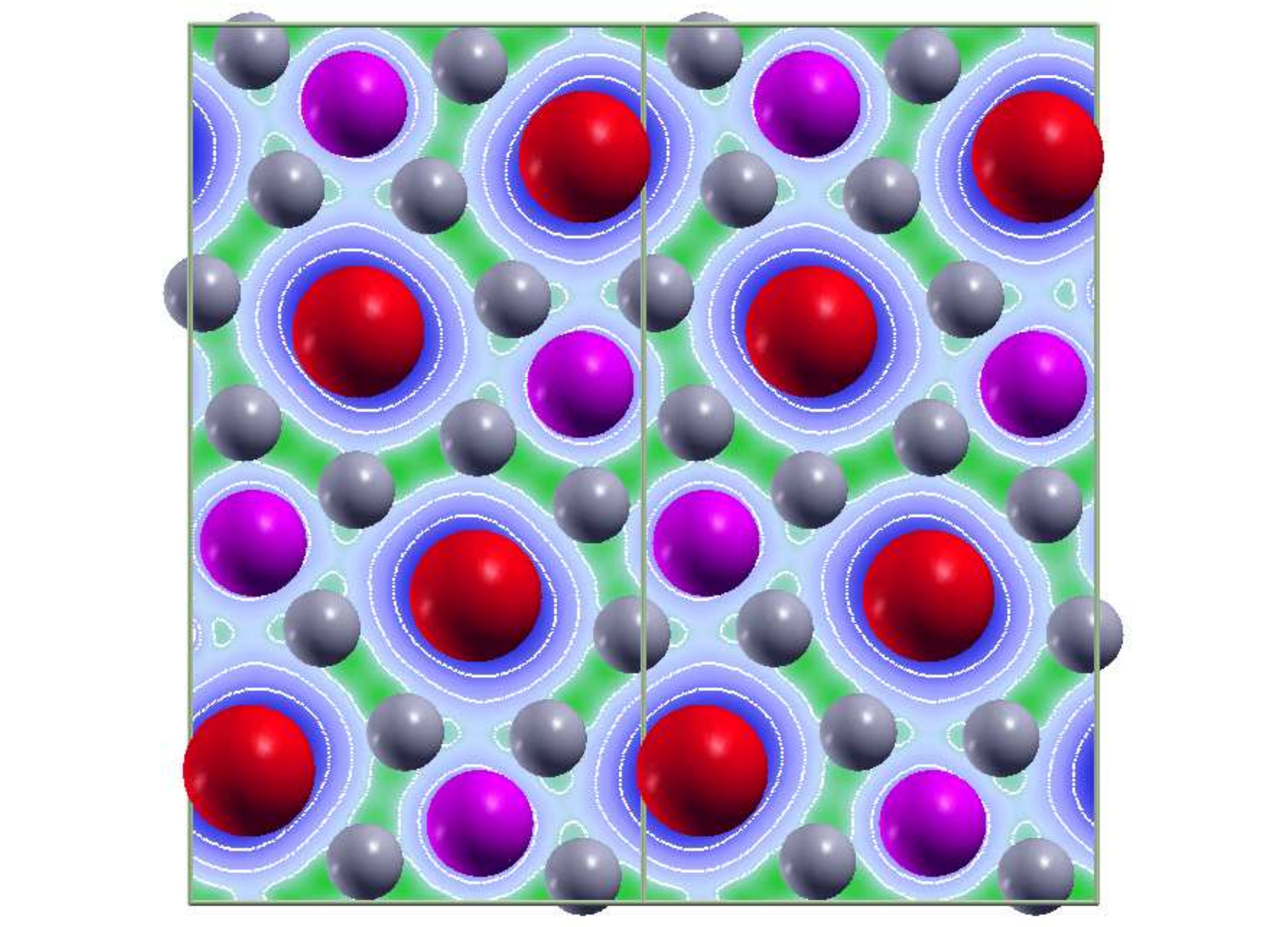} \\
\\
(c) \yb \\
\includegraphics[width=1\linewidth,angle=0]{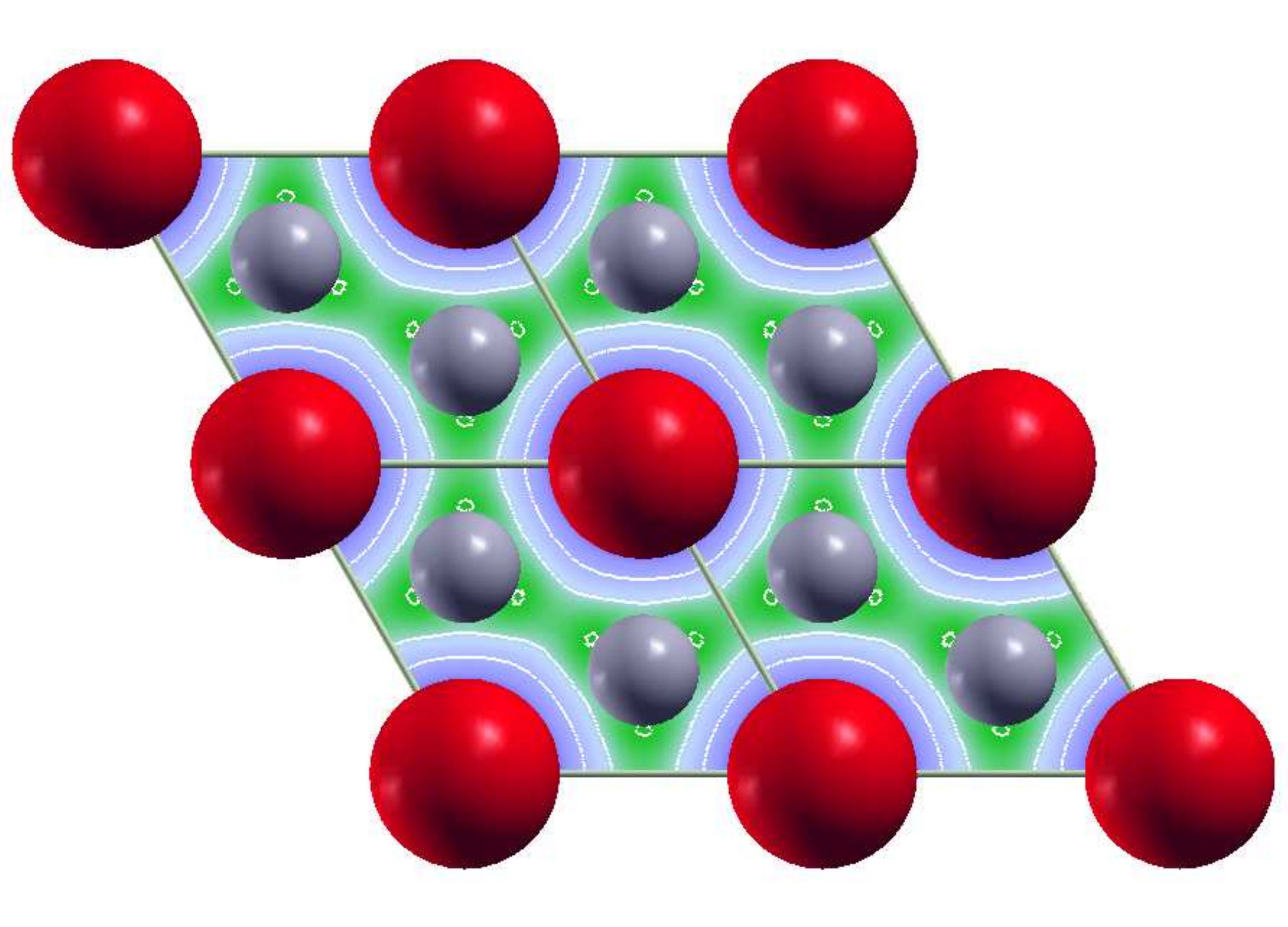}\\
\end{tabular}
\caption{\label{fig:DensContoursPlane} (Color online) Charge density contours in the boron plane on a linear scale for (a) \ybal, (b) \aybal~and (c) \yb.
For \yb we have doubled the unit cell in the \textbf{a} and \textbf{b} directions to aid visualisation of the bonding. Note the different topology of the
strong B-B bonds in the $\alpha$ and $\beta$ types, both contrasting
with the regular, hexagonal topology of YbB$_2$.}
\end{figure}

\begin{center}
\textbf{VI. Charge Density Analysis}
\end{center}
In Fig.~\ref{fig:DensContoursPlane} we show contour plots of the valence charge density in the boron plane for \ybal, \aybal~and \yb. The dominant chemical bonding resides within this boron plane and is driven by the short boron to boron distances, all of which are less than 1.9~\AA. The boron  bonding of these structures is very stable. For the high symmetry case of \yb~shown in Fig.~\ref{fig:DensContoursPlane}(c) we see that the bonding between the boron sites in its 6-membered rings is the same for every B-B bond. This bonding is of a similar character to that in graphite. In contrast for \ybal~and \aybal~all B-B bonds are not equivalent. The 5-member and 7-member rings determined by the crystallographic positions do not give rise to a dominant ring structure in the bonding charge density. Specific bonds within these 5-member and 7-member rings are stronger than others. 

As shown in Fig.~\ref{fig:DensContoursPlane}(a) the dominant B-B bonds in \ybal~form a 2D network of large loops each of which contain two formula units. The boron network is well connected in both the \textbf{a}-axis and \textbf{b}-axis directions. In Fig.~\ref{fig:DensContoursPlane}(b) \aybal~shows a different bonding network. The dominant B-B bonds form chains along the \textbf{a}-axis with double-boron branches. The difference in bonding anisotropy should also be reflected in anisotropy in transport behavior. As a result the bonding structure and distribution of the valence electrons can not be approximated as conforming to the ring structure as is the case in \yb.

The bond charge isocontours indicate that the bond population of the dominant bonds is approximately double that of the weak bonds. Interestingly, in both \ybal~and \aybal, the weak B-B bonds always lie between a Yb and an Al ion. There is a direct correlation between bond length and bond population. The weak bonds are all $\approx$1.86~\AA~in length, while all other B-B bonds are $\leq$1.77~\AA. Therefore the subtle symmetry changes between these compounds are significant to the electronic structure.


The different valence bond networks demonstrated to exist in \ybal, \aybal~and \yb~are likely then to affect the magnetic coupling between Yb moments. This may impact the RKKY interaction in these systems which may be the driving force behind magnetic instabilities. The uneven bond population surrounding the Yb ion also makes its environment anisotropic. This may be a key factor in producing the quantum criticality and superconductivity in \ybal.
\\

\begin{center}
\textbf{VII. Conclusions}
\end{center}
We have obtained strong evidence for a large Kondo scale driven by strong interaction between the Yb \textit{f}-state and the itinerant states
in \ybal~and \aybal. This is evidenced by strongly spin split bands and significant hybridization between the \textit{f}-states and itinerant states near the Fermi level.

We present a case for a key role of the angle subtended by the Yb ion to the boron ring in producing the large Kondo scale and possibly the superconductivity of \ybal.
The introduction of the 7-member coordination of the Yb ion in \ybal~allows the angle of $\approx 48^{\circ}$ to be subtended to the boron ring. This angle brings the compound close to the optimal $\approx 52^{\circ}$, for the maximization of $\Delta / E_0$ which promotes its large Kondo scale while still allowing for a large hybridization of the 4$f$-hole with the conduction states. The importance of the angle subtended by the Yb ion to the boron ring suggests a significant scope for the use of experiments involving pressure tuning via \textit{uniaxial} pressure that may be able to replicate these conditions in other materials. The compound \yb~is a strong candidate for such measurements.
The in-plane bonding structure amongst the dominant itinerant electrons in the boron sheets has been found to differ significantly between \ybal~and \aybal.

\begin{center}
\textbf{VIII. Acknowledgments}
\end{center}
We acknowledge an ICAM-I2CAM Institute for Complex Adaptive Matter travel grant that stimulated this collaboration.
This project was supported partially by DOE grant DE-FG02-04ER46111 and benefitted from interactions within
the Predictive Capability for Strongly Correlated Systems team of the Computational
Materials Science Network.

\bibliography{main}

\end{document}